\documentclass[conference]{IEEEtran}
\IEEEoverridecommandlockouts

\usepackage{cite}
\usepackage{amsmath,amssymb}
\usepackage{graphicx}
\usepackage{booktabs}
\usepackage{xcolor}
\usepackage{tikz}
\usetikzlibrary{arrows.meta,calc,fit,backgrounds}
\usepackage[hidelinks]{hyperref}

\hypersetup{
  pdftitle={The Content Leash: Mitigating Degenerate Solutions in Invertible Speaker Anonymization},
  pdfauthor={Nirjon Mobarak Talukder, Rafid Ahmed, Mohammad Ishrak Abedin, Abu Raihan Mostofa Kamal},
  pdfkeywords={speaker anonymization, voice privacy, invertible neural networks, knowledge distillation}
}

\def\BibTeX{{\rm B\kern-.05em{\sc i\kern-.025em b}\kern-.08em T\kern-.1667em\lower.7ex\hbox{E}\kern-.125emX}}

\definecolor{procedge}{HTML}{B5651D}
\definecolor{procfill}{HTML}{FDF1E4}
\definecolor{dataedge}{HTML}{37474F}
\definecolor{newedge}{HTML}{1E5AA8}
\definecolor{newfill}{HTML}{E8F0FB}
\definecolor{gapedge}{HTML}{C62828}
\definecolor{gapfill}{HTML}{FDECEC}
\definecolor{flowc}{HTML}{263238}
\tikzset{
  obj/.style={
    font=\small,
    >={Stealth[length=2.2mm,width=1.6mm]},
    proc/.style={rectangle, rounded corners=2pt, draw=procedge, fill=procfill,
                 line width=0.7pt, align=center, inner xsep=4pt, inner ysep=3pt,
                 minimum width=22mm, minimum height=9mm},
    data/.style={rectangle, rounded corners=2pt, draw=dataedge, fill=white,
                 line width=0.6pt, align=center, inner xsep=4pt, inner ysep=3pt,
                 minimum width=12mm, minimum height=8mm},
    focus/.style={data, line width=1.6pt},
    loss/.style={rectangle, rounded corners=2pt, draw=gray!65, fill=gray!10,
                 line width=0.6pt, align=center, inner sep=4pt, font=\footnotesize,
                 minimum width=46mm},
    newc/.style={rectangle, rounded corners=2pt, draw=newedge, fill=newfill,
                 line width=1pt, align=center, inner sep=4pt, font=\footnotesize},
    gapc/.style={rectangle, rounded corners=2pt, draw=gapedge, fill=gapfill,
                 line width=1pt, align=center, inner sep=4pt, font=\footnotesize},
    flow/.style={->, line width=0.8pt, draw=flowc},
    fix/.style={->, line width=1.3pt, draw=newedge},
    gaparr/.style={->, line width=1pt, draw=gapedge, dashed},
    lbl/.style={font=\scriptsize, text=flowc},
  }
}

\begin{document}

\title{The Content Leash: Mitigating Degenerate Solutions\\in Invertible Speaker Anonymization}

\author{
\IEEEauthorblockN{Nirjon Mobarak Talukder, Rafid Ahmed,\\
Mohammad Ishrak Abedin, and Abu Raihan Mostofa Kamal}
\IEEEauthorblockA{Department of Computer Science and Engineering\\
Islamic University of Technology, Dhaka, Bangladesh\\
Email: \texttt{\{nirjonmobarak, rafidahmed, ishrakabedin, raihan.kamal\}@iut-dhaka.edu}}
}

\maketitle

\begin{abstract}
Voice anonymization aims to protect a speaker's identity while preserving the linguistic
content of speech, enabling voice data to be shared and processed with reduced privacy
risk. This work focuses on reversible voice anonymization, where the original voice can be
recovered by an authorized user rather than being permanently discarded. However, existing
approaches can achieve apparent speaker anonymization by degrading the speech itself,
while their effectiveness may also depend on the speaker-verification representation used
during training. This creates a need for methods that protect speaker identity without
sacrificing speech intelligibility and that remain effective against unseen speaker
verifiers. We introduce content-preserving training constraints to improve the
privacy--utility trade-off by preventing anonymization from being achieved primarily
through speech degradation. We further introduce a verifier-generalization strategy that
incorporates a stronger speaker-verification representation during training, improving
anonymization robustness beyond the verifier used in the original training setup. The
resulting approach maintains intelligible speech, strengthens anonymization, and preserves
reliable voice restoration, with evaluation on unseen speakers and held-out speaker
verification models. On 38 unseen speakers, the approach raises the attacker's equal
error rate from a measured floor of 2.8\% to 26.3\%, and to 27.9\% against a held-out
verifier, with 11.8\% word error rate and 94.9\% speaker similarity on key-based
restoration. These improvements advance reversible voice anonymization toward practical
privacy-preserving speech applications, where anonymized speech can be safely processed
or shared while the original voice remains recoverable under authorized conditions.
\end{abstract}


\section{Introduction}

Speech carries the identity of the speaker as well as the words. Any system that stores
or transmits recordings therefore stores a biometric identifier. Speaker anonymization
removes that identifier while keeping the linguistic content usable, and it has become
an active research area through the VoicePrivacy
challenges~\cite{tomashenko2022vpc2020,tomashenko2022vpc2022plan}.

Most published systems are one-way. Once a recording is anonymized the original voice is
gone. This blocks legitimate uses such as a medical review, a legal dispute or a fraud
investigation, where an authorized party must be able to recover the original speaker.
Rano~\cite{wang2025rano} addresses this with a conditional invertible neural network
(cINN): the same network anonymizes in the forward direction and restores in the inverse
direction, conditioned on a pseudo-speaker embedding derived from a secret key.

This paper reports a paper-guided re-implementation of Rano and three modifications
motivated by our experimental observations. Our contributions are:

\begin{itemize}
\item We observe a degenerate solution in our re-implementation of the published training
      objective, where the triplet loss can be satisfied through severe spectrogram
      degradation rather than meaningful speaker transformation. A controlled experiment
      demonstrates this behavior under the evaluated configuration.
\item We add two constraints on the anonymized spectrogram, an anchor loss and a range
      loss, that mitigate the observed shortcut and substantially improve speech
      intelligibility in our experiments.
\item We observe that an anonymizer trained against a lightweight mel-domain verifier
      does not fool an independent verifier, and reduce this gap by distilling
      ECAPA-TDNN into a mel-domain verifier.
\item We report every metric against a measured floor, add a held-out verifier, seed the
      pseudo-speaker keys and report variance, and evaluate only on speakers absent from
      all training data.
\end{itemize}

\section{Related Work}

\textbf{Speaker anonymization.} The dominant family disentangles speech into speaker and
content representations, replaces the speaker representation, and resynthesizes. The
VoicePrivacy baseline replaces an x-vector~\cite{snyder2018xvector} with an average of
dissimilar vectors~\cite{srivastava2020design}; later work samples pseudo-speakers so that
the distribution of anonymized vectors is preserved~\cite{turner2020dpxvector}, uses
adversarial training to remove residual speaker information~\cite{perero2022autoencoder},
or carries content in phonetic posteriorgrams~\cite{sun2016ppg}. Signal-processing
methods such as McAdams-coefficient shifting~\cite{patino2021mcadams} need no training
and serve as a challenge baseline. All of these are one-way: the original voice cannot be
recovered.

\textbf{Invertible networks.} Normalizing flows build bijective mappings from affine
coupling layers~\cite{dinh2017realnvp,kingma2018glow}, and conditional variants make the
mapping depend on side information~\cite{ardizzone2019cinn}. Rano~\cite{wang2025rano}
applies a cINN to anonymization so that the forward and inverse directions share one
parameter set, which makes restoration exact rather than learned. Our work starts from
that design and examines what its training objective does and does not constrain.

\textbf{Distillation.} Knowledge distillation transfers the behavior of a trained network
to another~\cite{hinton2015distilling}; similarity-preserving distillation matches the
pairwise similarity structure of a batch rather than individual
outputs~\cite{tung2019similarity}. We use both terms to move a waveform-domain speaker
verifier into the mel domain.

\section{Background}

\subsection{Restorable anonymization with a cINN}

Let $X \in \mathbb{R}^{80 \times T}$ be the log-mel spectrogram of an utterance and
$f_\theta(\cdot\,;\cdot)$ a conditional invertible network. A key $k \sim \mathcal{N}(0,I)$
is mapped by an anonymization condition generator (ACG) to a pseudo-speaker embedding
$c$. The forward pass produces the anonymized spectrogram $X_a = f_\theta(X; c)$, and the
inverse pass with the same key recovers $X_r = f_\theta^{-1}(X_a; c) = X$ up to
floating-point precision. Restoration is therefore a property of the architecture rather
than something the network learns, which is what separates this family from one-way
anonymization. We measure a round-trip mean squared error of about $10^{-6}$ on the
spectrogram.

\subsection{The published objective}

Training uses two losses. The consistency loss requires that conditioning on the
speaker's own embedding leaves the spectrogram unchanged,
\begin{equation}
  \mathcal{L}_{\mathrm{cons}} = \mathrm{MSE}\big(X,\, f_\theta(X; s)\big),
  \label{eq:cons}
\end{equation}
where $s = V(X)$ is the embedding of the true speaker produced by a frozen speaker
verification network $V$, and $\mathrm{MSE}$ averages the squared error over all $80
\times T$ elements. The triplet loss drives anonymization,
\begin{equation}
  \mathcal{L}_{\mathrm{tri}} = \mathrm{ReLU}\!\big(m + d(a,c) - d(a,s)\big),
  \quad d(u,v) = 1 - \cos(u,v),
  \label{eq:tri}
\end{equation}
where $a = V(X_a)$ is the embedding of the anonymized spectrogram, $c$ is the
pseudo-speaker embedding, $m$ is a margin and $d$ is the cosine distance. The total
objective is $\mathcal{L} = \lambda_1 \mathcal{L}_{\mathrm{cons}} + \lambda_2
\mathcal{L}_{\mathrm{tri}}$.

\section{Method}

\subsection{The degenerate solution}

\begin{figure*}[t]
\centering
\resizebox{0.86\textwidth}{!}{%
\begin{tikzpicture}[obj]
\node[data] (X) at (0,0)      {$X$\\[-2pt]{\scriptsize original}};
\node[data] (s) at (1.4,3.5)  {$s=V(X)$\\[-2pt]{\scriptsize true speaker}};
\node[data] (c) at (1.4,-1.5) {$c$\\[-2pt]{\scriptsize condition}};

\node[proc] (fs)    at (3.4,2.0)  {cINN forward\\[-2pt]{\scriptsize same weights}};
\node[data] (xs)    at (6.6,2.0)  {$f_\theta(X;s)$};
\node[loss] (lcons) at (13.4,2.0)
  {$\mathcal{L}_{\mathrm{cons}} = \mathrm{MSE}(X, f_\theta(X;s))$\\[1pt]
   {\scriptsize acts on the pass conditioned on $s$; never sees $X_a$}};

\node[proc]  (fc)   at (3.4,0)  {cINN forward};
\node[focus] (xa)   at (6.6,0)  {$X_a$\\[-2pt]{\scriptsize anonymized}};
\node[proc]  (asv)  at (9.2,0)  {speaker verifier};
\node[loss]  (ltri) at (13.4,0)
  {$\mathcal{L}_{\mathrm{tri}}$ on $a=V(X_a)$\\[1pt]
   {\scriptsize toward $c$, away from $s$; acts on the embedding only}};

\draw[flow] (X) -- (fc);
\draw[flow] (X.north) |- (fs.west);
\draw[flow] (s.east) -| (fs.north);
\draw[flow] (c.east) -| (fc.south);
\draw[flow] (fs) -- (xs);
\draw[flow] (xs) -- (lcons);
\draw[flow] (fc) -- (xa);
\draw[flow] (xa) -- (asv);
\draw[flow] (asv) -- node[above, lbl] {$a$} (ltri);

\node[gapc] (gap) at (1.0,-3.1)
  {\textbf{the content of $X_a$ is unconstrained}\\[1pt]
   the triplet loss can be satisfied\\[-1pt]
   by \emph{degrading} the spectrogram};
\node[newc] (prop) at (7.6,-3.1)
  {\textbf{Proposed constraints on $X_a$}\\[2pt]
   $\mathcal{L}_{\mathrm{anchor}} = \mathrm{MSE}(X_a, X)$\;\;{\scriptsize limits content change}\\[1pt]
   $\mathcal{L}_{\mathrm{range}}$: keeps $X_a$ within $[X_{\mathrm{lo}}, X_{\mathrm{hi}}]$};

\draw[gaparr] (gap.north east) -- (xa.south west);
\draw[fix] (xa.south |- prop.north) -- (xa.south)
      node[midway, right, font=\scriptsize, text=newedge] {act on $X_a$ directly};
\end{tikzpicture}}
\caption{The training objective viewed as constraints on the anonymized spectrogram
$X_a$. The consistency loss acts on a second forward pass conditioned on the true
speaker embedding $s$ and never sees $X_a$. The triplet loss acts only on the speaker
embedding of $X_a$. Neither penalizes changes to its content (red). The two proposed
losses act on $X_a$ directly (blue).}
\label{fig:gap}
\end{figure*}
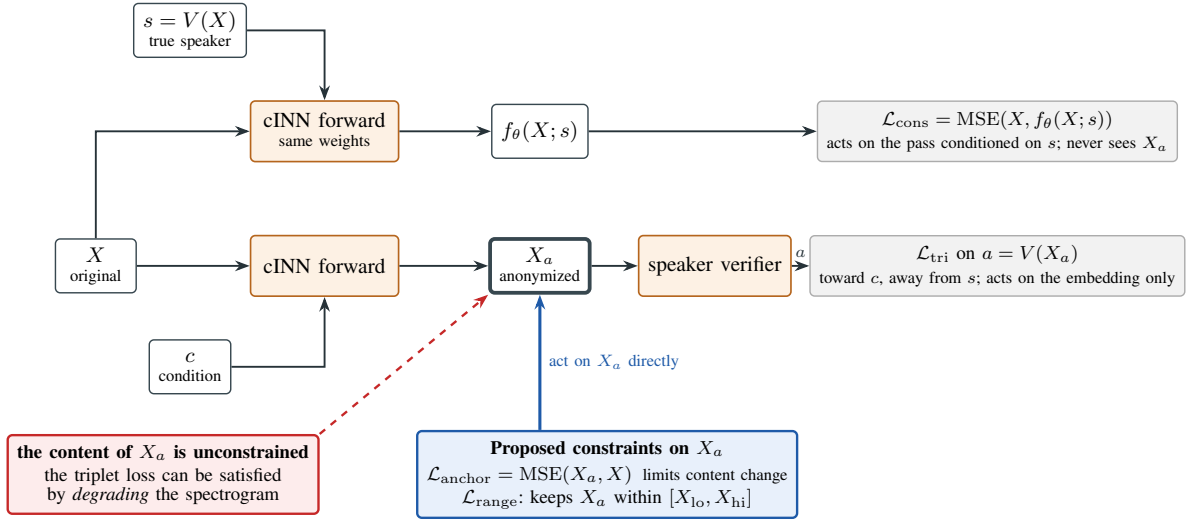

Equation~\eqref{eq:cons} constrains $f_\theta(X;s)$, a separate forward pass conditioned
on the true speaker. Equation~\eqref{eq:tri} constrains $a$, the embedding of the
anonymized spectrogram. Neither penalizes changes to $X_a$ beyond its speaker embedding,
as illustrated in Fig.~\ref{fig:gap}.

This creates a potential shortcut: the anonymization objective does not directly constrain
the content of $X_a$. In our controlled experiment, optimization converged to a degenerate
solution in which the triplet loss decreased despite severe speech degradation. The
consistency loss remained low throughout, because it constrains a pass conditioned on the
true speaker embedding rather than the anonymized output itself.

\subsection{Modification 1: content-preserving losses}

\textbf{Anchor loss.} The mean squared error between the anonymized and the original
spectrogram,
\begin{equation}
  \mathcal{L}_{\mathrm{anchor}} = \mathrm{MSE}\big(X_a,\, X\big),
  \label{eq:anchor}
\end{equation}
pulls $X_a$ back toward $X$. Linguistic content lives in the spectrogram, so bounding the
distance limits how much the content can change. This sets up a deliberate tension: the
triplet loss pushes the spectrogram to change and the anchor pulls it back, and the
equilibrium is the privacy--utility operating point, selectable through
$\lambda_{\mathrm{anchor}}$.

\textbf{Range loss.} A hinge on the dynamic range of the original utterance,
\begin{equation}
  \mathcal{L}_{\mathrm{range}} = \mathrm{mean}\big(
     \mathrm{ReLU}(X_a - X_{\mathrm{hi}}) + \mathrm{ReLU}(X_{\mathrm{lo}} - X_a)\big),
  \label{eq:range}
\end{equation}
where $X_{\mathrm{lo}}$ and $X_{\mathrm{hi}}$ are the per-utterance minimum and maximum
of $X$. The term is exactly zero while $X_a$ stays inside the envelope of the original
and grows linearly outside it. It is not redundant with the anchor: the anchor penalizes
all deviation, including the moderate movement that anonymization needs, whereas the
range loss is indifferent inside the envelope and penalizes only excursions outside it.
Values far outside the envelope are out of distribution for the vocoder; in the worst run
we observed, spectrogram values above $88$ produced non-finite audio in the
spectrogram-inversion path.

The complete objective is
\begin{equation}
\begin{aligned}
  \mathcal{L} = {}& \lambda_1 \mathcal{L}_{\mathrm{cons}}
              + \lambda_2 \mathcal{L}_{\mathrm{tri}}\\
              &+ \lambda_{\mathrm{anchor}} \mathcal{L}_{\mathrm{anchor}}
              + \lambda_{\mathrm{range}} \mathcal{L}_{\mathrm{range}}
              + \lambda_{\mathrm{det}} \mathcal{L}_{\mathrm{det}} ,
\end{aligned}
  \label{eq:total}
\end{equation}
where $\mathcal{L}_{\mathrm{det}} = \mathrm{mean}\big[(\log|\det J| / n)^2\big]$ is a
log-determinant regularizer, $J$ is the Jacobian of the forward pass, $n$ is the number
of elements it is summed over, and $\lambda_{\mathrm{det}} = 0.01$. It penalizes both
expansion and contraction of volume, which keeps the inverse pass numerically
well conditioned.

\subsection{Modification 2: an ECAPA-distilled training verifier}

An anonymizer can only learn to hide what its training verifier can see. Following the
original design, our re-implementation first used a lightweight mel-domain verifier
$V_{\mathrm{light}}$, a time-delay encoder trained on the same corpora. A model trained
against it drove the cosine similarity between anonymized and original embeddings in its
own space to $-0.09$, that is, past orthogonal, while an independent
ECAPA-TDNN~\cite{desplanques2020ecapa} still linked the same audio to the true speaker at
18.35\% equal error rate.

Training directly against ECAPA-TDNN is not practical: it consumes waveforms while the
anonymizer produces spectrograms, so a vocoder would be required inside every training
step. We therefore distill ECAPA-TDNN into a mel-domain verifier $V_{\mathrm{mel}}$. The
same speech crop is given to both networks in their own input format, ECAPA-TDNN stays
frozen, and $V_{\mathrm{mel}}$ is trained with
\begin{equation}
  \mathcal{L}_{\mathrm{distill}}
  = \frac1N\sum_{i=1}^{N}\big[1-\cos(g(h_i),e_i)\big]
  + \frac{\lambda_{\mathrm{rel}}}{N^{2}}\big\lVert HH^{\top}-EE^{\top}\big\rVert_F^{2},
  \label{eq:distill}
\end{equation}
where $N$ is the batch size, $e_i$ is the ECAPA-TDNN embedding, $h_i$ is the
$V_{\mathrm{mel}}$ embedding, both of unit length, $g$ is a linear read-out that matches
the two dimensionalities, and $H$ and $E$ stack the embeddings of the batch as rows. The
first term aligns each utterance individually. The second term, which follows
similarity-preserving distillation~\cite{tung2019similarity}, matches the pairwise cosine
similarities inside $V_{\mathrm{mel}}$'s own 256-dimensional space. We expect the second
term to matter most, because the triplet loss later operates on $h_i$ while $g$ is
discarded after distillation; we did not ablate the two terms separately.

Replacing the training verifier requires retraining the ACG, because it generates
conditions inside the verifier's embedding space. Pairing a new verifier with an old ACG
makes the triplet loss compare incompatible spaces.

\subsection{Modification 3: evaluation methodology}

Four protocol choices are applied throughout. Every metric is reported beside a measured
floor, obtained by running the same metric on unmodified speech, because an equal error
rate is uninterpretable without knowing how the verifier performs on clean audio. A
held-out verifier from a different architecture family, never used in any training stage,
is reported alongside the primary verifier. The pseudo-speaker keys are drawn from a
seeded generator and results are reported as a mean and standard deviation over three
seeds; unseeded keys hid a two-point spread in our early results. Finally, evaluation
uses only speakers absent from every training corpus, verified programmatically.

\section{Experimental Setup}

\textbf{Data.} The verifier, the ACG and the anonymizer are trained on
VCTK~\cite{vctk2019} and the \texttt{train-clean-100} subset of
LibriTTS~\cite{zen2019libritts}, split 85/15 at utterance level; the held-out 15\% is
used only for validation and checkpoint selection. Evaluation uses 303 utterances from 38
speakers of LibriTTS \texttt{test-clean}, verified disjoint from all training data. Up to
eight utterances per speaker were drawn at random with a fixed seed, keeping only
utterances with a reference transcript and speakers with at least four such utterances.
The vocoder used for the main results is fine-tuned on VCTK only, so no component of that
system has processed the evaluation speakers during training.

\textbf{Model.} The anonymizer is a cINN with 12 conditional coupling blocks (26.5\,M
parameters) operating on 80-band log-mel spectrograms at 22.05\,kHz with a hop of 256 and
a window of 1024. The ACG has 8 coupling blocks. $V_{\mathrm{mel}}$ and
$V_{\mathrm{light}}$ share the same architecture, a time-delay network with attentive
statistics pooling~\cite{okabe2018attentive} producing 256-dimensional embeddings.
Waveforms are synthesized with HiFi-GAN~\cite{kong2020hifigan}, fine-tuned for 30k steps
from public universal weights so that it matches the anonymizer's mel configuration.

\textbf{Training.} Distillation runs for 30k steps (AdamW~\cite{loshchilov2019adamw},
learning rate $10^{-3}$, batch 64, $\lambda_{\mathrm{rel}}=1$), the ACG for 80k steps,
and the anonymizer for 52k steps from random initialization with
$\lambda_{\mathrm{anchor}}=0.2$, $\lambda_2=5$, followed by 12k warm-started steps with
the final weights $\lambda_1=1$, $\lambda_2=3$, $\lambda_{\mathrm{anchor}}=1.0$,
$\lambda_{\mathrm{range}}=1.0$, $\lambda_{\mathrm{det}}=0.01$ and $m=0.3$. The reported
system is the checkpoint at step 8k of the second stage. Training used one RTX 6000 Ada;
evaluation and vocoder fine-tuning used one RTX 4060.

\textbf{Metrics and attack model.} Privacy is the equal error rate (EER) of cosine
scoring with ECAPA-TDNN. The attacker enrolls on original speech and is unaware that the
test speech has been anonymized, which corresponds to the \emph{ignorant} attacker of the
VoicePrivacy challenges~\cite{tomashenko2022vpc2022plan}. Every anonymized utterance is
scored against every original utterance, including the anonymized and original versions
of the same utterance, giving 2{,}417 genuine and 89{,}392 impostor trials; the floor
repeats the procedure with original speech on both sides and excludes the self-pairs. One
key is drawn per speaker and used for all of that speaker's utterances. A ResNet verifier
from SpeechBrain~\cite{ravanelli2021speechbrain} trained on VoxCeleb~\cite{nagrani2017voxceleb}
is held out and never used in training. Intelligibility is the word error rate (WER) of
Whisper large-v3~\cite{radford2023whisper}. Restoration quality is
$\mathrm{Sim}_{\mathrm{spk}}$, the cosine similarity between ECAPA-TDNN embeddings of the
original and the restored waveform.

\section{Results}

\subsection{Evidence of a degenerate solution}

Table~\ref{tab:ablation} is the central experiment. Both arms start from the same
checkpoint and train for 3{,}000 further steps on identical data with identical batch
size; the only difference is whether the two proposed losses are active.

\begin{table}[t]
\centering
\caption{Ablation of the two proposed losses. Both arms share initialization, data, batch
size and step count, and are evaluated on a 50-utterance subset of the evaluation set
(10 speakers, one key seed), where the ASR floor is 2.01\% WER. Real log-mel values on
this data reach about 10.2. Both arms use the earlier vocoder (see Section~\ref{sec:lim}).}
\label{tab:ablation}
\begin{tabular}{lccccc}
\toprule
 & EER & WER & mel & max & triplet \\
 & (\%) & (\%) & MSE & mel & loss \\
\midrule
Losses off & 41.60 & 117.58 & 6.551 & 12.4 & 0.0039 \\
Losses on  & 21.20 & \textbf{8.29} & 1.338 & 8.5 & 0.163 \\
Starting point & --- & --- & 1.294 & 9.2 & --- \\
\bottomrule
\end{tabular}
\end{table}

Removing the two losses makes the model look better on the privacy metric: 41.60\% EER,
approaching the 47.8\% reported in the original work. It reaches that value while the
speech is destroyed, at 117.58\% WER, which exceeds 100\% because the recognizer
hallucinates insertions into what is effectively noise. The mechanism is visible in the
statistics: five times the spectrogram distortion of the control, and a maximum value of
12.4 where real speech on this data reaches about 10.2. The triplet loss meanwhile
collapses to 0.0039, so by the published objective this model is excellent.

In our implementation, degrading the spectrogram was the cheapest way for the optimizer
to satisfy the triplet term, so the objective as realized here did not merely permit
content degradation but favored it. We note two limits on this observation. No official
implementation of the original method is available, so differences in architecture,
initialization or training schedule between our re-implementation and the original cannot
be ruled out as contributing factors. The word error rate of 11.9\% reported in the
original work also indicates that the released model does not exhibit this behavior. We
therefore present this as a property we observed in the evaluated configuration, not as a
defect of the published method.

\subsection{Verifier transfer}

Holding $\lambda_{\mathrm{anchor}}=0.2$ and $\lambda_2=5$ fixed and changing only the
training verifier, $V_{\mathrm{light}}$ yields 18.35\% EER at 8.71\% WER while the
ECAPA-distilled $V_{\mathrm{mel}}$ yields 39.58\% EER at 39.52\% WER. The privacy gain of
21.2 points is therefore not free at this anchor weight: intelligibility degrades
sharply, which is why the reported system raises $\lambda_{\mathrm{anchor}}$ to 1.0 and
trades part of that gain back for usable speech. Both runs were measured on an earlier
protocol that used seen speakers and a Griffin-Lim vocoder, so the absolute values are not
comparable with Table~\ref{tab:main}; the comparison between them is matched.

On the evaluation speakers, which no training stage saw, $V_{\mathrm{mel}}$ reaches a
cosine alignment of $0.730 \pm 0.098$ with ECAPA-TDNN, against $0.878$ measured on
training batches at the end of distillation. The pairwise similarity matrices of the two
verifiers correlate at $0.766$ on the same utterances, with a mean absolute difference of
$0.085$. This is consistent with findings on the transferability of attacks between a
surrogate and a target model~\cite{papernot2017practical,demontis2019why}.

\subsection{Main results}

\begin{table}[t]
\centering
\caption{Main results on 38 unseen speakers and 303 utterances. Floors are measured on
clean audio. EER is the mean and standard deviation of three seeded runs.}
\label{tab:main}
\begin{tabular}{lcccc}
\toprule
Metric & Floor & Ours & McAdams & Rano~\cite{wang2025rano} \\
\midrule
EER, ECAPA (\%)        & 2.83 & $\mathbf{26.31 \pm 0.25}$ & 8.44 & 47.81 \\
EER, ResNet (\%)       & 2.08 & \textbf{27.86}            & 6.37 & --- \\
WER (\%)               & 2.48 & \textbf{11.75}            & 3.61 & 11.91 \\
$\mathrm{Sim}_{\mathrm{spk}}$ (\%) & 94.90 & \textbf{94.90} & n/a & 87.75 \\
\bottomrule
\end{tabular}
\par\vspace{2pt}\parbox{\linewidth}{\footnotesize
  Rano~\cite{wang2025rano} values are taken from the original publication and were not
  reproduced under this protocol; its own reported floors are 3.20\% EER and 6.58\% WER.
  The $\mathrm{Sim}_{\mathrm{spk}}$ floor is the similarity of a vocoder round trip on
  unmodified speech, the ceiling any restoration can reach.}
\end{table}

Table~\ref{tab:main} gives the final system against the McAdams
baseline~\cite{patino2021mcadams} and the figures reported for Rano. The system raises
the attacker's error rate from 2.83\% to 26.31\%, a factor of 9.3, at 11.75\% WER against
a floor of 2.48\%. Restoration recovers 94.90\% speaker similarity with fully synthesized
phase, which equals the measured floor of a vocoder round trip on unmodified speech: the
inverse pass itself costs nothing measurable, and restoration quality is bounded by the
vocoder alone.

Two properties are worth separating. Against the McAdams baseline the trade is explicit:
roughly 8 points of word error rate spent for roughly 18 points of equal error rate, plus
invertibility, which that baseline cannot provide at any operating point. Relative to the
published Rano results, our system reports a lower ECAPA-TDNN EER and a higher restoration
similarity. The two WER values are close in absolute terms, but each should be read
against its own floor: Rano reports 11.91\% against a floor of 6.58\%, and we report
11.75\% against 2.48\%, so the cost of anonymization is larger in our system. Because Rano
was independently re-implemented and the evaluation protocols differ, these published
values are not a controlled baseline comparison. The matched ablation in
Table~\ref{tab:ablation} provides the direct evidence for the effect of the proposed
constraints.

The held-out ResNet reports 27.86\% against ECAPA-TDNN's 26.31\%, and it is the stronger
verifier of the two, with the lower clean floor. Neither the speakers nor this verifier
were seen during training, so the privacy result is not an artifact of optimizing against
the evaluation verifier.

\subsection{Key security}

Table~\ref{tab:key} measures restoration with an incorrect key at a controlled distance
$D_{\mathrm{key}} = \lVert c - c' \rVert_2$ between the correct and the incorrect
condition. With any incorrect key the recovered similarity falls to about 4\%. For
reference, the mean ECAPA-TDNN similarity between utterances of different speakers on
this evaluation set is $7.66\%$ (sd $8.98$), so an incorrect key leaves the restored
audio no closer to the original speaker than an unrelated recording.

\begin{table}[t]
\centering
\caption{Restoration quality with an incorrect key, over 303 utterances. For incorrect
keys, lower is better. A flat response denies an attacker a search gradient. These runs
use the earlier vocoder (see Section~\ref{sec:lim}).}
\label{tab:key}
\begin{tabular}{lccc}
\toprule
Case & mean $D_{\mathrm{key}}$ & Ours (\%) & Rano~\cite{wang2025rano} (\%) \\
\midrule
Correct key             & 0.00 & 95.28 & 87.75 \\
$0.1<D_{\mathrm{key}}<0.2$ & 0.14 & \textbf{4.09} & 12.78 \\
$0.2<D_{\mathrm{key}}<0.3$ & 0.24 & \textbf{4.41} & 9.49 \\
$0.3<D_{\mathrm{key}}<0.4$ & 0.35 & \textbf{4.30} & 4.66 \\
\bottomrule
\end{tabular}
\par\vspace{2pt}\parbox{\linewidth}{\footnotesize
  The two distance measures are not the same quantity and the rows are aligned only by
  bin label: we use the Euclidean distance between unit-norm \emph{conditions}, whereas
  \cite{wang2025rano} uses the cosine distance between \emph{keys} sampled from Gaussian
  noise. The Rano column is shown for reference, and the comparison should be read as a
  trend rather than a matched measurement.}
\end{table}

The flat response matters more than the absolute values. If restoration quality improved
as a guessed key approached the correct one, an attacker could try a key, measure how
speaker-like the output is, and refine. The reported figures for the original system rise
from 4.66\% to 12.78\% as the wrong key gets closer, which gives such a gradient. Our
system stays near 4.2\% at every distance we tested, so a failed attempt reveals nothing
about how close it was.

\section{Limitations}
\label{sec:lim}

Our equal error rate is bounded by distillation fidelity. $V_{\mathrm{mel}}$ reproduces
ECAPA-TDNN only approximately, at a cosine alignment of $0.730$ on held-out speakers, and
anonymization pressure beyond that point optimizes
into the region where the two disagree, which buys no measurable privacy while still
costing intelligibility. The evaluation covers 38 speakers from a single corpus and one
attack model, an ignorant attacker; informed attackers retrained on anonymized speech are
not tested. We report EER, WER and restoration similarity, but not the voice
distinctiveness and pitch correlation used in the VoicePrivacy challenges, so our results
are not directly comparable with challenge submissions. Restoration is evaluated on the
anonymized spectrogram rather than on the anonymized audio, so a deployment would need
lossless storage of that representation. The ablation removes both proposed losses
together, uses a single seed on a 50-utterance subset, and therefore does not separate
their individual contributions.

The results also mix two vocoders. Tables~\ref{tab:ablation} and~\ref{tab:key} were
measured with a HiFi-GAN fine-tuned on VCTK and LibriTTS \texttt{test-clean}, whose
fine-tuning data therefore included the evaluation utterances, while
Table~\ref{tab:main} uses a vocoder fine-tuned on VCTK alone. Repeating the main
evaluation with each vocoder changed the equal error rate by 0.43 points and the
restoration similarity by 0.38 points, both inside the seed-to-seed spread, so the two
sets of numbers are comparable; we report the figures separately rather than merging
them.

\section{Conclusion}

In our paper-guided re-implementation, the published training objective exhibited a
degenerate solution in which a high speaker-verification error rate coincided with
severely degraded intelligibility: 41.6\% EER at 117.6\% WER. An anchor loss and a range
loss on the anonymized spectrogram mitigated this behavior in the evaluated
configuration, bringing word error rate to 8.3\% under matched conditions. Distilling a
waveform-domain verifier into the mel domain reduced a second gap, between fooling the
training verifier and fooling an independent one, and was worth 21 points of equal error
rate at matched settings, at a cost in intelligibility that the anchor weight then trades
back. The resulting system reaches $26.3 \pm 0.3$\% equal error rate against a floor of
2.8\%, with 11.8\% word error rate and 94.9\% restoration similarity, and shows a flat
response of about 4\% to incorrect keys at every distance tested.


\bibliographystyle{IEEEtran}
\bibliography{refs}

\end{document}